\documentclass[%
 reprint,
 superscriptaddress,
 graphicx,
 amsmath,amssymb,
 aps,prr,
 longbibliography,
]{revtex4-2}

\usepackage{graphicx}% Include figure files
\usepackage{dcolumn}% Align table columns on decimal point
\usepackage{bm}% bold math
\usepackage[breaklinks=true,colorlinks=true,linkcolor=blue,urlcolor=blue,citecolor=blue]{hyperref}
\usepackage{upgreek}

\begin{document}

\title{Radiation pressure acceleration of high-quality ion beams using ultrashort laser pulses}%\thanks{A footnote to the article title}%
%alternative titles
% Generation of high-quality ion beams from radiation pressure acceleration of thin targets
% High-quality radiation pressure ion acceleration from thin targets via the mitigation of electron heating
\author{H.-G. Jason Chou}
\email[]{jasonhc@slac.stanford.edu} 
\affiliation{%
 High Energy Density Science Division, SLAC National Accelerator Laboratory, Menlo Park, CA 94025, USA
}% 
\affiliation{Department of Physics, Stanford University, Stanford, CA 94305, USA}
\author{A. Grassi}
\affiliation{%
 High Energy Density Science Division, SLAC National Accelerator Laboratory, Menlo Park, CA 94025, USA
}%
\author{S. H. Glenzer}
\affiliation{%
 High Energy Density Science Division, SLAC National Accelerator Laboratory, Menlo Park, CA 94025, USA
}%
\author{F. Fiuza}
\email[]{fiuza@slac.stanford.edu}
\affiliation{%
 High Energy Density Science Division, SLAC National Accelerator Laboratory, Menlo Park, CA 94025, USA
}

\begin{abstract}
The generation of compact, high-energy ion beams is one of the most promising applications of intense laser-matter interactions, but the control of the beam spectral quality remains an outstanding challenge. We show that in radiation pressure acceleration of a thin solid target the onset of electron heating is determined by the growth of the Rayleigh-Taylor-like instability at the front surface and must be controlled to produce ion beams with high spectral quality in the light sail regime. The growth rate of the instability imposes an upper limit on the laser pulse duration and intensity to achieve high spectral beam quality and we demonstrate that under this optimal regime, the maximum peak ion beam energy per nucleon is independent of target density, composition, and laser energy (transverse spot size). Our predictions are validated by two- and three-dimensional particle-in-cell simulations, which indicate that for recent and upcoming experimental facilities using ultrashort ($\lesssim 25$ fs) laser pulses it is possible to produce $100 - 300$ MeV proton beams with $\sim 30\%$ energy spread and high laser-to-proton energy conversion efficiency.
\end{abstract}

\maketitle

%%%%%%%%%%%%%%%%%%%%%%%%%%%%%%%%%%
High-energy, high spectral quality (quasi-monoenergetic) ion beams are important for a variety of applications, including radiography of high-energy-density materials and plasmas \cite{borg2002,Rygg2008ProtonImplosions.}, isochoric heating of materials \cite{Patel2003IsochoricBeam}, fast ignition of inertial confinement fusion targets \cite{roth2001}, injectors for conventional accelerators \cite{Antici2008NumericalSource}, and tumor therapy \cite{Bulanov2008AcceleratingPulses,Kraft2010Dose-dependentBeams}. Intense laser-matter interactions have long been seen as a very promising route to drive compact ion beam sources. However, despite significant progress over the last two decades, the control of the ion beam spectral quality remains an outstanding challenge. 

Among different laser-driven acceleration mechanisms explored, Radiation Pressure Acceleration (RPA) \cite{Wilks1992AbsorptionPulses,Esirkepov2004HighlyRegime,Macchi2005LaserPlasmas,Robinson2008RadiationPulses}--- in particular, the Light Sail (LS) regime using very thin (sub-micron) targets --- has attracted significant interest due to the possibility to produce narrow energy spread ion beams with high laser-to-ion energy conversion efficiency. Nevertheless, the experimental characterization of this acceleration scheme and observation of narrow energy spread ion beams have been elusive \cite{Henig2009Radiation-PressurePulses,Kar2012IonPressure,Steinke2013}. A critical challenge has been the requirement of low electron heating for efficient momentum transfer from the laser to the ions, and to avoid other competing ion acceleration mechanisms to develop and dominate. 

Over the last years, there has been a focus on the use of increasingly higher laser intensities to produce high ion energies. Indeed, recent experiments using intensities $> 10^{20}$ W/cm$^2$ and thin ($< 1$ $\upmu$m) targets reached near 100 MeV maximum proton energies \cite{Kim2016RadiationPulses,Wagner2016MaximumTargets,Higginson2018Near-100Scheme}. However, the measured energy spectra were very broad, with only a small number of ions at this high energy, likely due to the contribution of large sheath electric fields arising from strong electron heating \cite{Shen2021ScalingFoils}. While it is well established that the use of a Circularly Polarized (CP) laser at near normal incidence to the target surface can help mitigate electron heating \cite{Macchi2005LaserPlasmas} by both $\mathbf{J}\times\mathbf{B}$ \cite{Kruer1985JxBLight,May2011} and Brunel mechanisms \cite{Brunel1987Not-so-resonantAbsorption}, the development of corrugations at the laser-target interaction surface due to different instabilities \cite{Sentoku2000MagneticPlasmas,Palmer2012Rayleigh-TaylorLaser,Eliasson2015InstabilityLaser,Sgattoni2015Laser-drivenStructures,Gode2017RelativisticInteractions,Wan2020EffectsAcceleration} and finite laser spot size effects \cite{Klimo2008MonoenergeticPulses,Dollar2012FiniteTargets} can contribute to strong electron heating even when a CP laser is employed. It remains unclear which electron heating mechanism most strongly impacts the spectral quality of the accelerated ions and what the best route is to produce ion beams with high energy and high spectral quality. 

%%%%%%%%%%%%%%%%%%%%%%%%%%%%%%%%%%
In this Letter, we show that the growth of the Rayleigh-Taylor-like Instability (RTI) at the front surface of thin targets is the dominant electron heating mechanism and leads to a significant increase of the energy spread of accelerated ion beams in the LS regime. The growth of the RTI imposes an upper limit on the laser pulse duration and intensity required to produce ion beams with peaked (quasi-monoenergetic) spectra, and we demonstrate that under this optimal regime, the maximum ion beam peak energy per nucleon is independent of target density, composition, and laser energy (transverse spot size). These predictions are validated by two- (2D) and three-dimensional (3D) Particle-In-Cell (PIC) simulations over a wide range of laser and target conditions. These findings have important implications for the optimization of ion acceleration from laser-solid interactions and to fully harness the potential of RPA for the generation of compact, high-quality ion beams for a wide range of applications.

In the LS regime the acceleration experienced by a thin target due to the radiation pressure of an intense laser can be approximated by $a_{\rm RPA}\simeq 2v_\text{HB}^2/l_0$ in the non-relativistic limit, where $l_0$ is the target thickness and $v_\text{HB}=\sqrt{(1+R)I/(2m_in_ic^3)}$ is known as the hole boring velocity, with $R\leqslant1$ the laser reflection coefficient, $I$ the laser intensity, $m_i$ the ion mass, $n_i=Zn_0$ the ion density (with $n_0$ the initial target density $n_0$ and $Z$ the ion charge number), $c$ the speed of light in vacuum. For negligible electron heating $R\simeq1$ a quasi-monoenergetic ion beam can be produced and an analytical solution to the final peak energy per nucleon $\epsilon_0$ can be obtained for a constant laser intensity \cite{Macchi2009LightReexamined}: $\epsilon_0=m_pc^2\xi^2/[2(\xi+1)]$, where $\xi=c a_0^2\tau_0 m_en_c/(m_in_il_0)$, with $a_0 \simeq 0.85\sqrt{I [\text{W/cm}^{2}](\lambda_0[\mu\text{m}])^2/10^{18}}$ the laser peak normalized vector potential, $\tau_0$ the laser pulse duration, $m_e$ the electron mass, and $n_c=m_e\omega_0^2/(4\pi e^2)$ the critical density associated with laser propagation in the plasma, where $e$ is the elementary charge, $\omega_0$ and $\lambda_0$ the laser frequency and wavelength respectively.

The efficiency of RPA is significantly reduced when there is strong electron heating, \emph{i.e.} $R \ll 1$. This has motivated the interest in using a CP laser at near normal incidence in order to suppress electron heating \cite{Macchi2005LaserPlasmas}. However, even for such a configuration electron heating can still arise (\emph{e.g.} \cite{Klimo2008MonoenergeticPulses, Wan2020EffectsAcceleration}). The causes and consequences of electron heating with a CP laser are not fully understood and are crucial for the optimization of RPA.

To study the onset of electron heating and its impact on the quality of LS accelerated ion beams we have performed a series of 2D and 3D PIC simulations with the code OSIRIS \cite{Fonseca2002,Fonseca2008One-to-oneSimulations}. We model a CP laser irradiating a planar target at near normal incidence. The laser is launched in the longitudinal $x_1$ direction. The typical size of the simulation box in 2D (3D) simulations is 400 (300) $c/\omega_0$ longitudinally and $250\,c/\omega_0$ transversely in $x_2$ (and $x_3$). The 2D (3D) simulations use 16 (8) particles per cell per species and a spatial resolution of 0.2 (0.5) $c/\omega_{pe}$ in each direction, where $\omega_{pe}=\sqrt{4\pi e^2n_0/m_e}$ is the electron plasma frequency. The time step is chosen according to the Courant-Friedrichs-Lewy condition and we have used a third order particle interpolation scheme for improved numerical accuracy.

    \begin{figure}[!htb]
    \includegraphics[width=.45\textwidth]{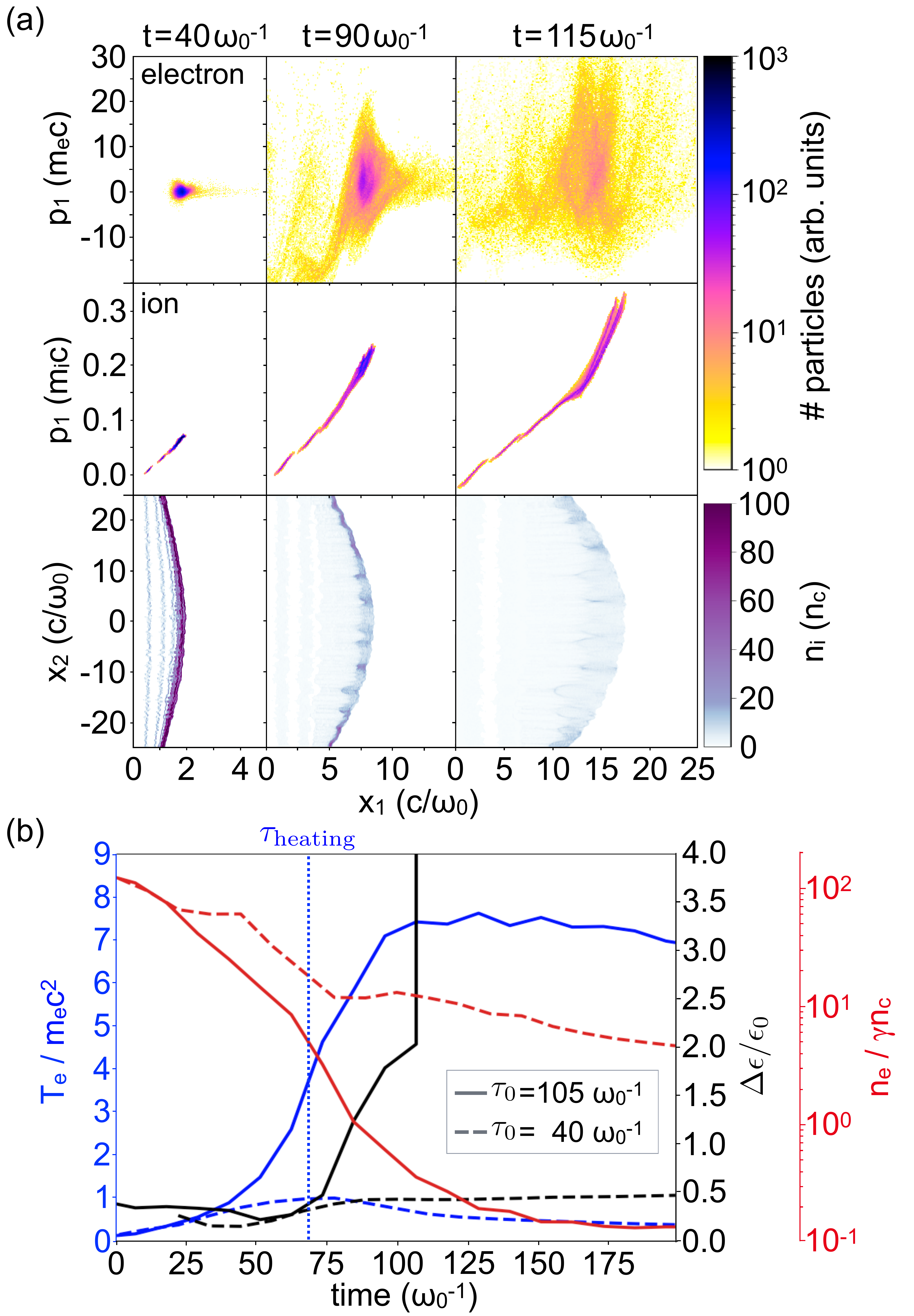}
    \caption{\label{fig1} Results of 2D PIC simulation of the interaction of an intense Gaussian CP laser pulse with a thin target. (a) Longitudinal (top) electron and (middle) ion phase spaces and (bottom) ion density profile for a laser with pulse duration $\tau_0 = 105\,\omega_0^{-1}$. (b) Temporal evolution of $T_e$ (blue, left axis), ion beam energy spread $\Delta\epsilon/\epsilon_0$ (black, right black axis) within a $10^\circ$ opening angle from the laser propagation direction, and $n_e/(\gamma n_c)$ (red, rightmost axis). The time $t = 0$ is defined as $\tau_0/2$ before the laser peak intensity reaches the target.}
    \end{figure}

Typical results are illustrated in Fig.~\ref{fig1}(a) for a laser with $a_0=15$, Gaussian longitudinal and transverse profiles with Full-Width-Half-Maximum (FWHM) of intensity duration $\tau_0 = 105\,\omega_0^{-1}$ and focal spot (at $1/e^2$ beam width) $w_0 = 50\,c/\omega_0$ irradiating a target with density $n_0 = 250n_c$ and thickness $l_0 = 0.085\,c/\omega_0$ (corresponding to $\sim 0.014$ $\upmu$m for $\lambda_0 = 1$ $\upmu$m). We observe that indeed even with CP there is the onset of strong electron heating after a relatively short interaction time. At $t = 90\,\omega_0^{-1}$ the electron and ion phase spaces show that electrons have been heated to relativistic temperatures, which resulted in a strong space-charge field, the rapid expansion of the target, and the continued broadening of the ion energy spread. By $t = 115\,\omega_0^{-1}$ the target expansion leads to the onset of relativistic transparency, RPA is terminated, and the ion energy distribution is no longer peaked. These are also illustrated in the temporal evolution of the electron temperature $T_e$ (average kinetic energy of electrons), FWHM ion energy spread $\Delta\epsilon/\epsilon_0$, and $n_e/(\gamma n_c)$ (the ratio of the electron density, $n_e$, and relativistic critical density at the target front surface, where $\gamma$ is the average Lorentz factor of the electrons), as shown in Fig.~\ref{fig1}(b). Given the general interest in highly directional beams for most applications, we consider the ion beam spectrum within a $10^\circ$ opening angle from the laser propagation direction. The simulation results reveal that the ion beam energy spread increases sharply following the rapid growth of $T_e$. We define this time, $\tau_\text{heating}$, as the time for which the rate of increase of the electron temperature, $dT_e/dt$, is maximum.

We find that the onset of strong electron heating is related to the emergence of large transverse density modulations in the target [Fig.~\ref{fig1}(a)]. The penetration of the laser in the lower density regions associated with these modulations triggers significant electron heating (see Appendix~\ref{appA}) with the temperature reached being comparable to that observed in simulations with a linearly polarized laser (not shown here). To confirm that this is playing a dominant effect in both the observed electron heating and increase in ion energy spread, we have repeated the same simulation but using a laser pulse with a duration $\tau_0 = 40\,\omega_0^{-1} < \tau_\text{heating}$. In this case, we observe no density modulations, the target remains relativistically overdense, and, indeed, both $T_e$ and $\Delta\epsilon/\epsilon_0$ remain low (see also Appendix~\ref{appA}). The ion beam energy spread saturates at $t \simeq 2 \tau_0$ and remains stable long after the laser-plasma interaction has finished [Fig.~\ref{fig1}(b)]. It is thus crucial to understand the origin of the density modulations and how their development depends on the laser and plasma conditions.

The growth of instabilities at the target surface from laser-plasma interactions has been recognized as an important source of transverse density modulations and there has been significant discussion on which instabilities are dominant, including the Weibel instability \cite{Sentoku2000MagneticPlasmas}, RTI \cite{Gamaly1993InstabilityBeam,Pegoraro2007PhotonPulse, Palmer2012Rayleigh-TaylorLaser,Khudik2014TheStability,Eliasson2015InstabilityLaser,Sgattoni2015Laser-drivenStructures}, and electron-ion coupling instabilities \cite{Wan2016PhysicalAcceleration,Wan2020EffectsAcceleration}. 

In order to study the mechanism responsible for the corrugations relevant for electron heating, and isolate the effects of surface instabilities, we have performed a parameter scan of 2D simulations with a long, plane-wave CP laser at normal incidence. We vary the laser $a_0 = 5 - 200$, target composition $1 \leqslant A/Z \leqslant 4$ (with $A$ the ion mass numbers), density ($n_0 = 40 - 500\,n_c$; covering the range from liquid hydrogen to solid-density targets for a laser wavelength of $1$ $\upmu$m), and thickness ($l_0 = 0.08 - 40\,c/\omega_0$). The target thickness is always $l_0 \geqslant l_\text{opt}$ to ensure stability of the target. Note that $l_\text{opt}= a_0\lambda_0n_c/(\sqrt{2}\pi n_0)$ \cite{Macchi2009LightReexamined} is the optimal target thickness for which the acceleration is maximized by minimizing the total target mass while guaranteeing that the target remains relativistically opaque. We analyze the growth of ion density modulations by computing the transverse Fourier modes ($k_{x_2}$) of the longitudinal ($x_1$) displacement of the relativistic critical surface ($n \simeq \gamma_0n_c$; where $\gamma_0=\sqrt{1+a_0^2/2}$ is the electron Lorentz factor), as a function of the transverse ($x_2$) position.

\begin{figure}[t]
    \includegraphics[width=.48\textwidth]{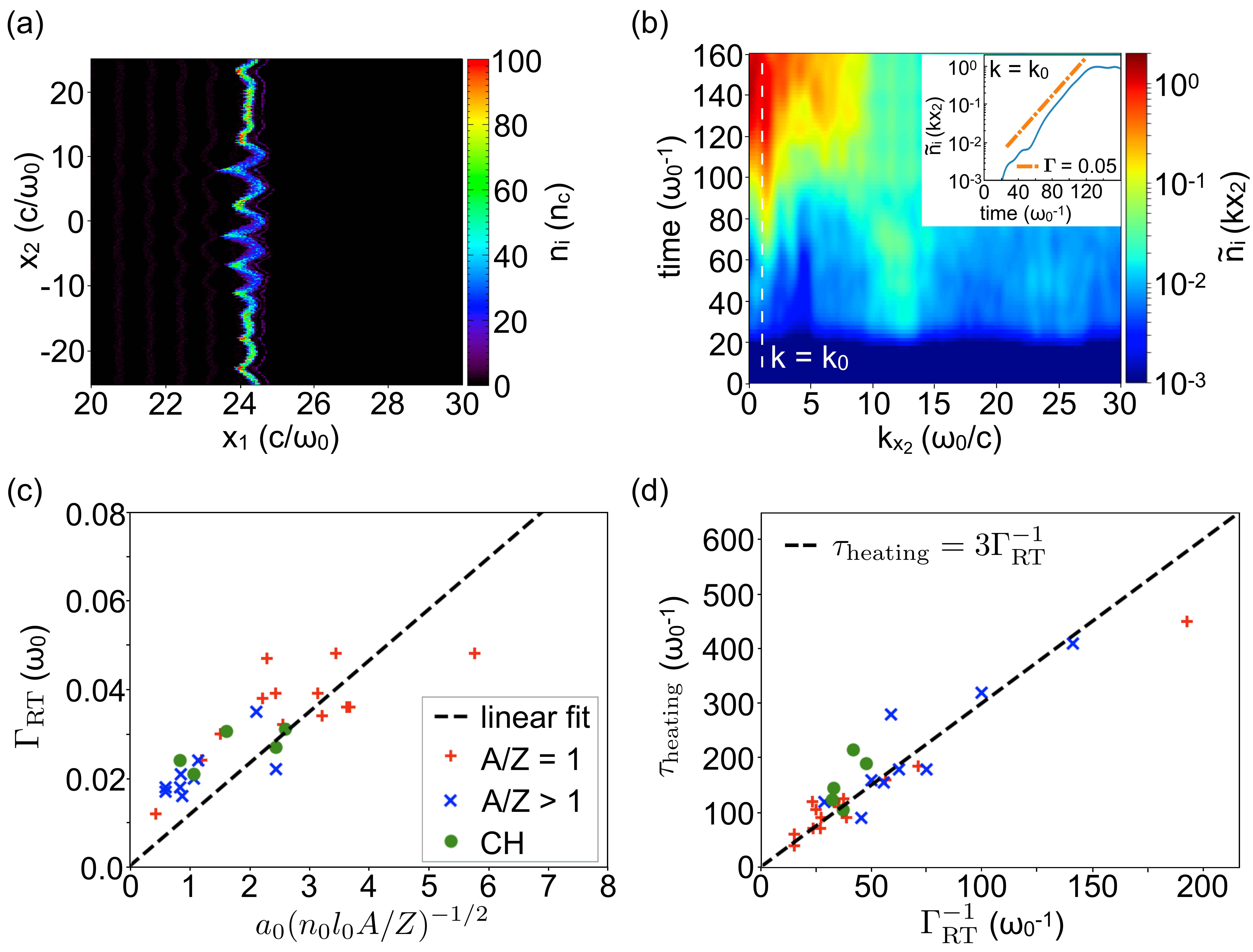}
    \caption{\label{fig2} (a) Development of surface corrugations in ion density at $t = 90\,\omega_0^{-1}$. (b) Evolution of Fourier modes at the relativistic critical surface of the corrugation amplitudes $\widetilde{n}_i(k_{x_2})$. The inset shows the growth rate of the $k_{x_2}=k_0$ mode and its linear fit. (c) Scaling of the measured growth rate of the $k_{x_2}=k_0$ mode of the surface corrugations. (d) Strong correlation between the electron heating time $\tau_\text{heating}$ and the growth time of the RTI. Colored symbols are measurements from 2D PIC simulations.}
    \end{figure}

Fig.~\ref{fig2}(a) and (b) illustrate the growth of different modes for a simulation with the same laser intensity and target parameters of Fig.~\ref{fig1}(a). Note that the target remains opaque to the laser during the time of the analysis and thus the measurements of the growth rate and saturation level are not affected by the onset of relativistic transparency. The fastest growing modes are observed at $k_{x_2}\gg k_0$, with $k_0=2\pi/\lambda_0$ [\emph{e.g.} $k_{x_2}\simeq 13\,\omega_0/c\,\equiv k_{max}$ in Fig.~\ref{fig2}(b) at $t\simeq 30\,\omega_0^{-1}$]. However, these modes saturate at relatively low amplitudes and do not lead to significant electron heating. The dominant density modulations are associated with the mode with $k_{x_2}\simeq k_0$ [Fig.~\ref{fig2}(b)] and we observe the onset of strong electron heating ($\tau_\text{heating} \simeq 85\,\omega_0^{-1}$) during the linear growth and saturation of this mode.
This suggests that the onset of strong electron heating is related to laser-driven RTI for which the dominant mode is $k_{x_2}\simeq k_0$ \cite{Eliasson2015InstabilityLaser,Sgattoni2015Laser-drivenStructures}. This is in contrast to recent work which argued that 
electron heating was associated with the non-linear stage of the
electron-ion coupling instabilities driven with $k_{x_2}\gg k_0$ \cite{Wan2020EffectsAcceleration}. To further confirm that the mode with $k_{x_2} = k_0$ is dominant in terms of electron heating, we have performed additional simulations with a smaller transverse domain ($\lesssim 0.3\,\lambda_0$), in which the $k_0$ mode is prohibited but high-$k$ modes are still present. We ran these simulations up to $t\simeq 3\,\tau_\text{heating}$ (with $\tau_\text{heating}$ measured from the simulation with the large transverse box size) and in all cases no significant electron heating or increase in ion energy spread are observed (see Appendix~\ref{appB}).

For the laser-driven RTI the growth rate is $\Gamma_\text{RT} \propto \sqrt{a_{\rm RPA}k_0}$ \cite{Eliasson2015InstabilityLaser,Sgattoni2015Laser-drivenStructures}. Linear fits to the measured growth rates of the $k_0$ mode over the large parameter scan of our study confirm the scaling [Fig.~\ref{fig2}(c); see also the example fit in Fig.~\ref{fig2}(b)]: $\Gamma_\text{RT}[\omega_0]\simeq 0.5\,a_0\left(n_0[n_c]l_0[c/\omega_0]Am_p/(Zm_e)\right)^{-1/2}$, where $m_p$ is the proton mass. Furthermore, we find a strong correlation between $\tau_\text{heating}$ and the growth time of the instability, with $\tau_\text{heating} \simeq 3\,\Gamma^{-1}_\text{RT}$, as shown in Fig.~\ref{fig2}(d). This indicates that, in order to suppress or significantly mitigate electron heating, the duration of the laser pulse $\tau_0$ should be shorter than a threshold value $\hat{\tau}_0 \equiv \tau_\text{heating}$, defining an optimal regime for LS acceleration based on the pulse duration as:

    \begin{eqnarray}
    \tau_0 \text{[fs]}\leqslant \hat{\tau}_0 \equiv
        350\, a_0^{-1}\left(l_0\,[\mathrm{\upmu m}]\lambda_0\,[\mathrm{\upmu m}]\frac{A}{Z}\frac{n_0}{n_c}\right)^{1/2}.
    \label{eqn:tausc}
    \end{eqnarray}

We note that surface corrugations can also arise due to the transverse variation of the laser intensity when a finite spot size is used, where the non-uniform acceleration across the surface would result in a change of the surface shape, triggering strong electron heating. Significant reshaping of the surface occurs when the displacement along the axis of the laser is comparable to the spot size $w_0$ \cite{Wan2020EffectsAcceleration}, from which we can similarly obtain a condition on the pulse duration: $\tau_0\leqslant 200\,a_0^{-1}\left(w_0[\mathrm{\upmu m}]l_0[\mathrm{\upmu m}]A n_0/(Z n_c)\right)^{1/2}$. Comparing this with Eq.~(\ref{eqn:tausc}) reveals that for relativistically opaque ($a_0 < n_0/n_c$) targets and for typically used spot sizes $w_0 \gtrsim$ 3 $\lambda_0$, the effect of RTI is dominant and thus imposes the main limitation on the pulse duration.

    \begin{figure}[!t]
    \includegraphics[width=.48\textwidth]{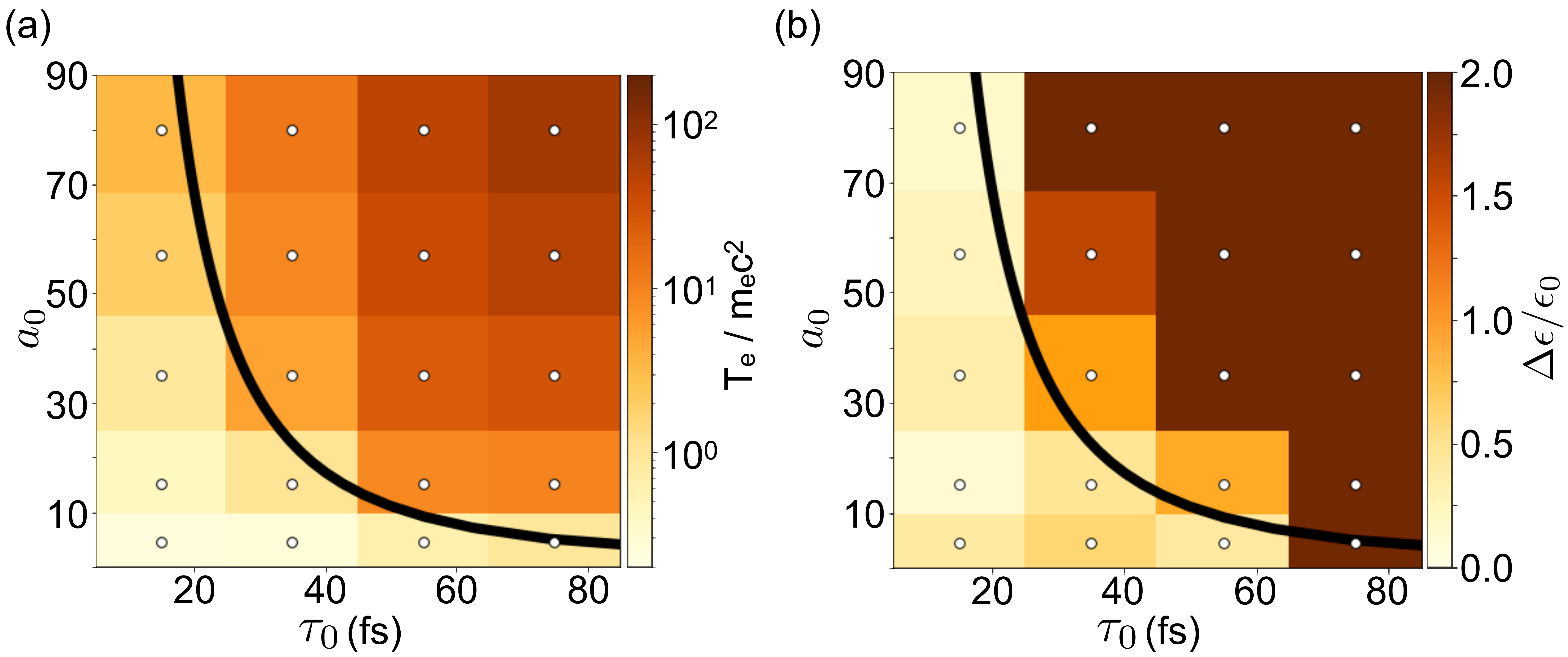}
    \caption{\label{fig:a0t0} (a) Electron temperature $T_e$ and (b) ion beam energy spread $\Delta\epsilon/\epsilon_0$ measured from 2D PIC simulations of a 1$\,\upmu$m wavelength
    Gaussian laser pulse with duration $\tau_0$ and spot size $w_0=7.6$ $\upmu$m interacting with a thin solid target with $n_0 = 250\,n_c$ and $l_0=l_\text{opt}$. $T_e$ is measured at the end of the laser interaction when the maximum is observed. $\Delta\epsilon/\epsilon_0$ is measured for protons within a $10 ^\circ$ opening angle at $t\simeq2\tau_0$. The black curve corresponds to the prediction of Eq.~\ref{eqn:tausc} and the white dots denote the parameters sampled by the simulations.}
    \end{figure}

Previous works have considered the role of RTI in the context of RPA \cite{Bulanov2010,Wang2021Laser-IonAccelerator,Yu2010StableFoil,Zhou2016ProtonRegime,Wu2014SuppressionPulses}, proposing advanced laser and target configurations, such as
curved \cite{Wang2021Laser-IonAccelerator} and mixed-species \cite{Yu2010StableFoil} targets or single-cycle \cite{Zhou2016ProtonRegime} and elliptically polarized \cite{Wu2014SuppressionPulses} lasers to mitigate RTI. These are either challenging to implement in practice \cite{Wang2021Laser-IonAccelerator,Zhou2016ProtonRegime} (and have not yet been proven to be effective experimentally), or still lead to significant electron heating \cite{Yu2010StableFoil,Wu2014SuppressionPulses}, which will negatively impact the ion beam spectral quality in the LS regime. More importantly, a quantitative understanding of the laser pulse duration and intensity required to mitigate RTI was not established and, as we show here, is critical to producing ion beams with high spectral quality.
 
In Fig.~\ref{fig:a0t0} we demonstrate that Eq.~(\ref{eqn:tausc}) is robust over a wide range of laser and target parameters even when realistic Gaussian transverse and temporal pulse profiles are considered. The prediction of Eq.~(\ref{eqn:tausc}) marks the transition from low-to-high electron heating [Fig.~\ref{fig:a0t0}(a)] and consequently from low-to-high energy spread of the accelerated ion beam [Fig.~\ref{fig:a0t0}(b)]. We have also repeated some of the simulations in the optimal pulse duration regime using a small, but finite laser incidence angle and confirm that these results are still valid. We have found that in general for an incidence angle $\lesssim$ 10$^\circ$, as typically used experimentally, electron heating is maintained at a low level and the quality of the ion beam remains similar to the case with normal incidence.

This model has important implications for the optimization of LS ion acceleration. When we consider the optimal target thickness for LS, $l_0 = l_\text{opt}$, Eq.~(\ref{eqn:tausc}) imposes an upper limit on $a_0$
\begin{equation}
        a_0 \lesssim \hat{a}_0 \equiv \frac{350^2}{\sqrt{2}\pi}\frac{A}{Z}\left(\frac{\lambda_0\,[\mathrm{\upmu m}]}{\tau_0\,\text{[fs]}}\right)^2, \label{eqn:a0}
\end{equation}   
which equivalently leads to a maximum peak energy per nucleon $\hat{\epsilon}_0$ for the accelerated ion beam
\begin{equation}
        \hat{\epsilon}_0 \equiv m_pc^2\frac{\hat{\xi}^2}{2(\hat{\xi}+1)}, \text{ where } \hat{\xi}\simeq 20 \frac{\lambda_0\,\text{[$\upmu$m]}}{\tau_0\,\text{[fs]}}.\label{eqn:eps0}
\end{equation}

Remarkably, for this high-quality (narrow energy spread) acceleration regime, the maximum peak energy per nucleon given by Eq.~(\ref{eqn:eps0}) is independent of target density, composition ($A/Z$), and laser energy (transverse spot size). 

\begin{figure}[!htb]
    \includegraphics[width=.47\textwidth]{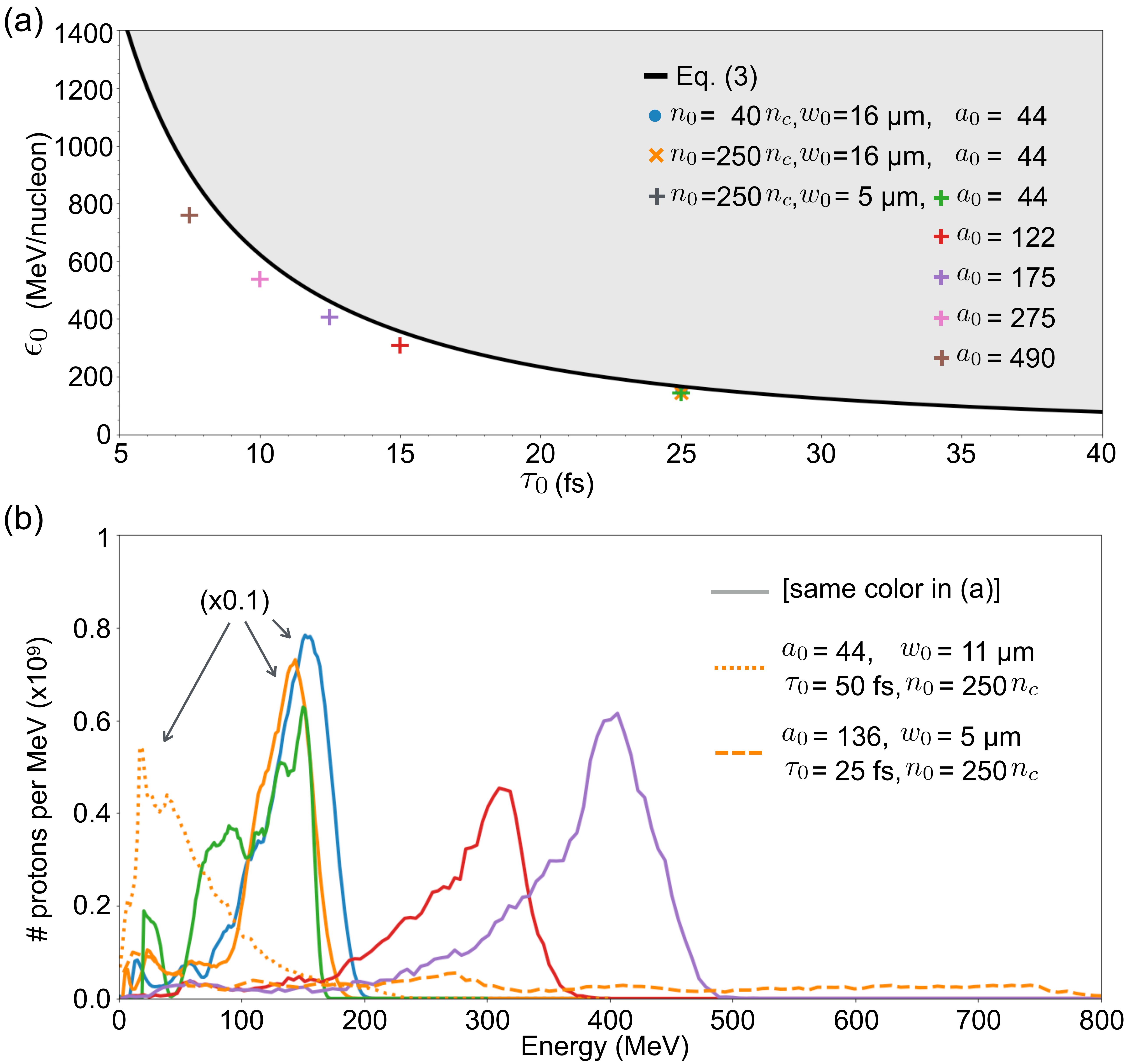}
    \caption{\label{fig4} (a) 
    Peak proton beam energy $\epsilon_0$ as a function of laser duration $\tau_0$ in high-quality regime of LS acceleration. Colored symbols are measurements from 3D PIC simulations, the solid line is the prediction of Eq. (\ref{eqn:eps0}), and the gray shaded area above the solid line indicates the parameter space outside the high-quality acceleration regime. (b) Proton spectra from 3D PIC simulations with $a_0=\hat{a}_0$ and $\tau_0=\hat{\tau}_0$ (solid lines), $a_0=\hat{a}_0$ and $\tau_0 = 2 \hat{\tau}_0$ (dotted), and $a_0 \simeq 3 \hat{a_0}$ and $\tau_0=\hat{\tau}_0$ (dashed). The spectra are measured within a $10^\circ$ opening angle from the laser propagation direction at $t\simeq 2\tau_0$.}
    \end{figure}

The prediction of Eq.~(\ref{eqn:eps0}) has been validated against a series of full 3D PIC simulations that use a Gaussian laser pulse profile. We have considered target densities of either $n_0 =$ 40 or 250$\,n_c$ with $l_0 = l_\text{opt}$. The pulse duration was varied in the range $\tau_0 = 7.5 - 25$ fs and for each duration the laser $a_0$ was chosen to be equal to the maximum $\hat{a}_{0}$ given by Eq.~(\ref{eqn:a0}). In all cases, we observe stable acceleration of the protons via LS leading to the generation of narrow energy spread proton beams with peak energies $\epsilon_0$ in very good agreement with the prediction of Eq.~(\ref{eqn:eps0}), as shown in Fig.~\ref{fig4}(a). For the case with $\tau_0 = 25$ fs, we have performed two additional simulations for different laser spot size (and thus, total energy) and target density. We observe that the peak proton energies obtained are very similar, confirming that indeed for the high-quality acceleration condition derived here the beam energy is independent of laser energy and target density. 

Fig.~\ref{fig4}(b) shows the resulting proton spectra and confirms the generation of high-quality proton beams when the conditions for $\tau_0$ and $a_0$ given by Eqs. (\ref{eqn:tausc})--(\ref{eqn:a0}) are satisfied. Under these optimal conditions, we observe the generation of peaked spectra with $\Delta \epsilon/\epsilon_0 \simeq 20 - 40$\% and with 3\% of the total laser energy being carried by this beam within a $10^\circ$ opening angle. The total laser-to-proton energy conversion efficiency into $4\pi$ is $20\%$. To further confirm the importance of our model, we have repeated the simulation with $\tau_0 = 25$ fs  but with either a higher intensity $a_0 = 136$ ($\simeq 3 \hat{a}_0$) or a longer pulse $\tau_0 = 50$ fs (= 2$\hat{\tau}_0$), while keeping the same laser pulse energy. In both cases we observe a very significant increase of the energy spread ($> 300\%$) and reduction of the coupling efficiency ($<1\%$).

%%%%%%%%%%%%%%%%%%%%%%%%%%%%%%%%%%
In summary, we have shown that electron heating arising from the growth of the RTI plays a dominant role in the spectral quality of ion beams accelerated in the interaction of an intense laser with thin solid- or liquid-density targets. To produce beams with high spectral quality it is critical to limit the laser pulse duration to the growth time of the RTI. This, in turn, also limits the maximum laser intensity that can be used and the maximum peak energy of the accelerated ion beam. For example, for 100 TW class lasers, the pulse duration should be limited to $\lesssim 60$ fs. The majority of previous laser-driven ion acceleration experiments in this regime have used linearly polarized lasers (see \cite{Zeil2010} for a summary), but the few experiments using short pulses and circular polarization have reported the generation of narrow energy spread ion beams \cite{Henig2009Radiation-PressurePulses,Steinke2013}, consistent with our model. For the parameters of high-power state-of-the-art and near future laser systems with $\tau_0\simeq$ 15 -- 25 fs, such as the ELI-NP \cite{Doria2020}, Apollon 10 PW \cite{papadopoulos}, and EP-OPAL \cite{EP-OPAL} facilities, full 3D PIC simulations demonstrate the possibility to produce $\sim 100 - 300$ MeV proton beams with $\sim 30\%$ energy spread and high coupling efficiency. These findings have important implications for the optimization of future experiments and to fully harness the promise of radiation pressure acceleration to produce high-quality ion beams for a wide range of applications.

%%%%%%%%%%%%%%%%%%%%%%%%%%%%%%%%%%
\begin{acknowledgments}
This work was supported by the U.S. Department of Energy SLAC Contract No. DEAC02-76SF00515, by the U.S. DOE Early Career Research Program under FWP 100331, and by the DOE FES under FWP 100182. The authors thank the OSIRIS Consortium, consisting of UCLA and IST (Portugal) for the use of the OSIRIS 4.0 framework. %and the visXD framework. 
Simulations were performed at Cori (NERSC) and Theta (ALCF) through ERCAP and ALCC computational grants.
\end{acknowledgments}

\appendix
%\counterwithin{figure}{section}

\section{}\label{appA}

\begin{figure}[!htb]
    \includegraphics[width=.45\textwidth]{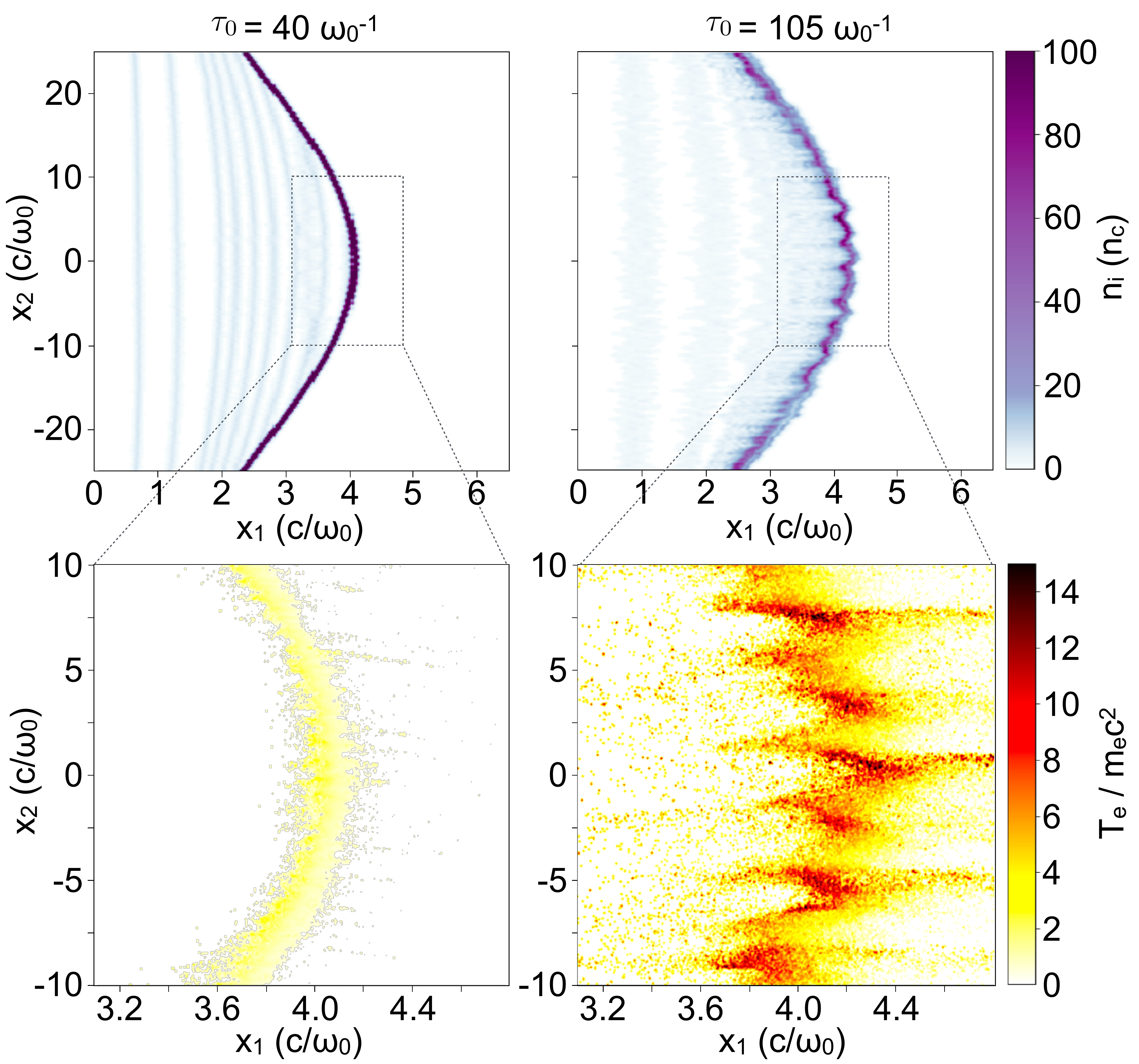}
    \caption{\label{fig5}Comparison between the short- (left) and long-pulse (right) simulations in Fig.~\ref{fig1}. Top: ion density; bottom: local electron temperature. All plots are shown at $t=70\,\omega_0^{-1}$, which is approximately both $\tau_\text{heating}$ for the long-pulse ($\tau_0 = 105\,\omega_0^{-1}$) case and the end of interaction time for the short-pulse ($\tau_0 = 40\,\omega_0^{-1}$) case. }
    \end{figure}

As surface modulations on the laser wavelength scale develop, the laser is able to penetrate these modulations giving rise to effective electron heating, for example via the Brunel mechanism. This is illustrated in Figure~\ref{fig5} for the same simulation parameters of Fig.~\ref{fig1}. In the long-pulse case (right column), the laser can penetrate the concave “valleys” in the surface density modulation, and accelerate electrons. The map of the local $T_e$ (average kinetic energy of electrons) shows indeed that the heating is happening at the walls of these concave “valleys” and consistent with direct acceleration by the laser electric field. The location of the hot spots of $T_e$ oscillates (from top to bottom of the valleys for instance) in accordance with the phase of the laser electric field. 

Note that the time shown in Fig.~\ref{fig5} corresponds to approximately the end of laser-plasma interactions for the short-pulse case of Fig.~\ref{fig1}(b). We can see that for short-pulse case (left column) the target remains stable, showing negligible surface rippling and very low electron heating. 

\section{}\label{appB}

\begin{figure}[!htb]
    \includegraphics[width=.4\textwidth]{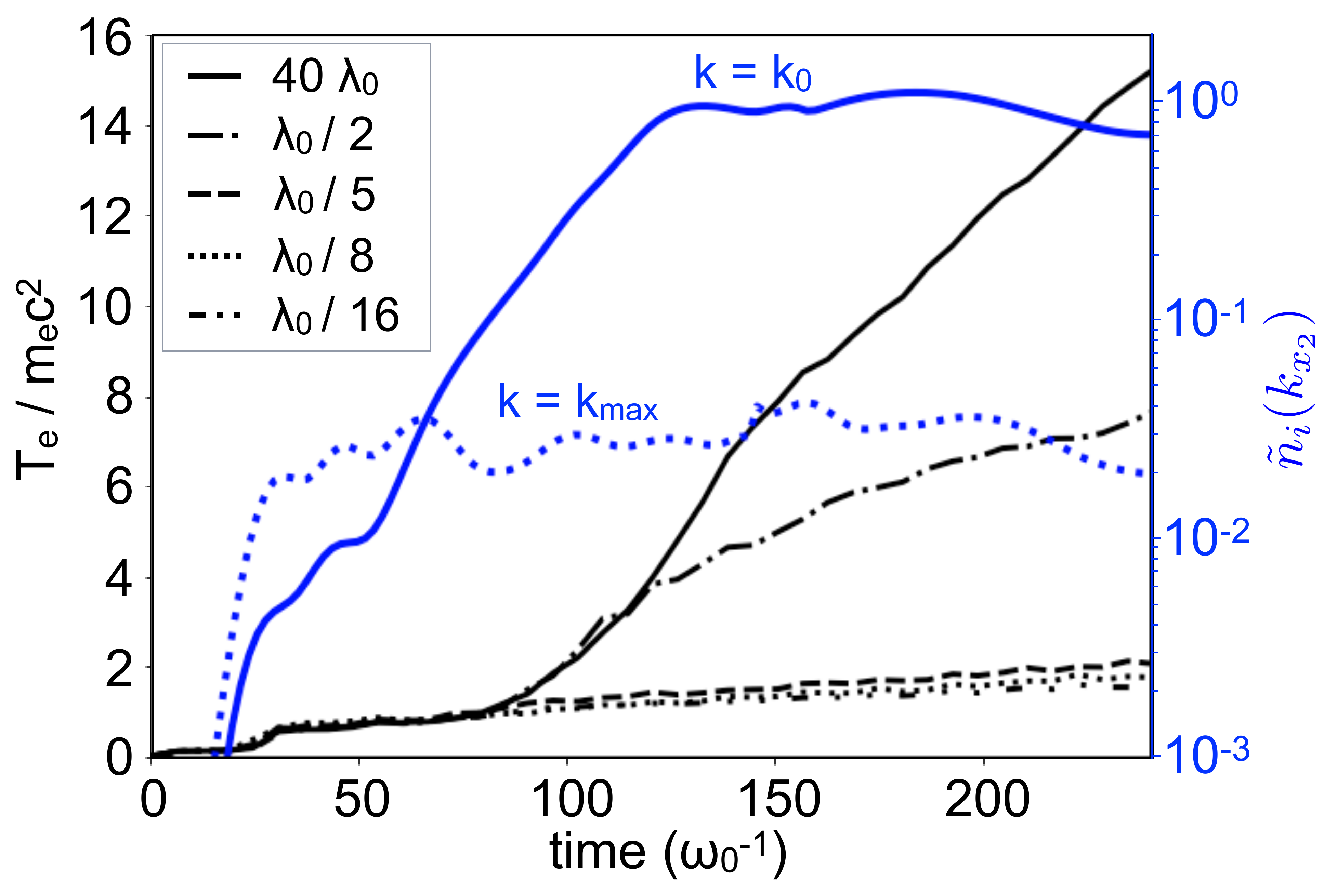}
    \caption{\label{fig6} Evolution of electron temperature for different transverse simulation domain sizes (black, left axis). The solid black curve corresponds to the simulation in Fig.~\ref{fig2}(a) and (b) with a transverse box size of $40\,\lambda_0$ and for this case the growth of the $k_{x_2} = k_0$ (solid) and $k_{x_2} = k_{max}$ (dotted) modes are shown (blue, right axis).}
    \end{figure}

Figure~\ref{fig6} shows the results of a set of 2D simulations with different transverse domains, where the same laser and target parameters as in Fig.~\ref{fig2}(a) and (b) were used. We observe that for transverse domain sizes $< \lambda_0/2$, where the dominant RTI mode is prohibited, no significant electron heating is observed. High-k modes can still grow, but $T_e$ remains very low. There are two distinct modes that grow and saturate with different growth rates and on different time scales:
the fastest growing mode $k_{x_2}\simeq13\,\omega_0/c\,\equiv k_{max}$, which is always captured and saturates at around $t\simeq50\,\omega_0^{-1}$,
and the RTI mode $k_{x_2}=k_0$, which saturates at $t > 120\,\omega_0^{-1}$.
Note that the $k_{max}$ mode saturates
much earlier than the observed $\tau_\text{heating}\simeq85\,\omega_0^{-1}$
, which is concurrent with the development of the RTI mode. In particular we see that strong electron heating starts near the saturation time of the RTI.

%%%%%%%%%%%%%%%%%%%%%%%%%%%%%%%%%%
%\bibliographystyle{apsrev4-2}
\bibliography{ref_manual_20220325}% Produces the bibliography via BibTeX.

\end{document}